\documentclass[aps,prb,nofootinbib,superscriptaddress,twocolumn]{revtex4}
\usepackage{graphicx}

\newcommand{\msr}{$\mu$SR}
\newcommand{\dBS}{Bi$_{2-x}$Fe$_x$Se$_3$}
\newcommand{\BS}{Bi$_2$Se$_3$}

\begin{document}

\title{The Nature of Magnetic Ordering in Magnetically Doped Topological Insulator Bi$_{2-x}$Fe$_x$Se$_3$}
\author{Z.~Salman}
\email{zaher.salman@psi.ch}
\affiliation{Laboratory for Muon Spin Spectroscopy, Paul Scherrer Institute, CH-5232 Villigen PSI, Switzerland}
\author{E. Pomjakushina}
\affiliation{Laboratory for Development and Methods, Paul Scherrer Institute, CH-5232 Villigen PSI, Switzerland}
\author{V. Pomjakushin}
\affiliation{Laboratory for Neutron Scattering, Paul Scherrer Institute, CH-5232 Villigen PSI, Switzerland}
\author{A. Kanigel}
\affiliation{Department of Physics, Technion - Israel Institute of Technology, Haifa 32000, Israel}
\author{K.~Chashka}
\affiliation{Department of Physics, Technion - Israel Institute of Technology, Haifa 32000, Israel}
\author{K. Conder}
\affiliation{Laboratory for Development and Methods, Paul Scherrer Institute, CH-5232 Villigen PSI, Switzerland}
\author{E. Morenzoni}
\affiliation{Laboatory for Muon Spin Spectroscopy, Paul Scherrer Institute, CH-5232 Villigen PSI, Switzerland}
\author{T. Prokscha}
\affiliation{Laboratory for Muon Spin Spectroscopy, Paul Scherrer Institute, CH-5232 Villigen PSI, Switzerland}
\author{K. Sedlak}
\affiliation{Laboratory for Muon Spin Spectroscopy, Paul Scherrer Institute, CH-5232 Villigen PSI, Switzerland}
\author{A. Suter}
\affiliation{Laboratory for Muon Spin Spectroscopy, Paul Scherrer Institute, CH-5232 Villigen PSI, Switzerland}

\date{\today}

\begin{abstract}
  We present a detailed investigation of the magnetic and structural
  properties of magnetically doped 3D topological insulator \BS. From
  muon spin relaxation measurements in zero magnetic field, we find
  that even 5\% Fe doping on the Bi site turns the full volume of the
  sample magnetic at temperatures as high as $\sim 250$K. This is also
  confirmed by magnetization measurements. Two magnetic ``phases'' are
  identified; the first is observed between $\sim 10-250$K while the
  second appears below $\sim 10$K. These cannot be attributed to
  impurity phases in the samples. We discuss the nature and details of
  the observed magnetism and its dependence on doping level.
\end{abstract}
\maketitle

\section{Introduction}
Topological insulators (TIs) are 3D materials with a band gap in the
bulk-electronic structure but with gapless, delocalized, surface
states\cite{Hasan10RMP}. The surface-states are protected by
time-reversal symmetry (TRS), and cannot be gapped without changing
substantially the bulk-electronic structure. This special relation
between the bulk and surface states is the basic difference between
these systems and systems showing ordinary surface states. The
protected surface states are believed to be robust to disorder,
interactions, and thermal fluctuations, potentially leading to
room-temperature device applications\cite{Moore09NP}. In recent years,
doped TIs have been investigated due to potential new phenomena when
the topological surface states interact with impurities or other
electronic states in the bulk. For example, it was found that by
intercalating a prototypical TI, \BS, with Cu it becomes
superconducting below $\sim 4$K \cite{Hor10PRL}. More recently, the
interaction between the surface states of a TI and magnetic impurities
have also been attracting attention since such impurities are expected
to break TRS. In this context, an important issue is the stability of
the TI surface states to the presence of impurities.

% It is predicted theoretically that the type of interaction between
% the magnetic impurities (RKKY) is dramatically altered by the
% metallic surface states, and its nature can be tuned, from
% ferromagnetic, antiferromagnetic and even frustration by varying the
% density of magnetic impurities \cite{Zhu11PRL}. The modification of
% the surfaces state by these impurities is of great importance from a
% fundamental point of view as well as potential applications in
% spintronic devices \cite{Murakami04PRL,Biswas10PRB}. An advantage of
% using doped TI to construct multi-layer structures is that high
% quality interfaces can be acheived due to the almost perfect
% structural matching between the different layers.

In the bulk, it was shown that a TI with magnetic impurities can have
long range magnetic order in the metallic \cite{Choi04PSS,Hor10PRB}
and insulating \cite{Yu10S} phases. On the surface, long range order
can also be formed independent of the bulk magnetic ordering, as the
RKKY interaction induced by the Dirac fermions is generally
ferromagnetic when the Fermi energy ($E_F$) is close to the Dirac
point.  Both effects can lead to breaking of TRS, resulting in a gap
opening at the Dirac point that makes the surface Dirac fermions
massive. Chen et al. \cite{Chen10S} have shown that the Dirac gap can
be observed in magnetically doped samples with or without bulk
ferromagnetism. Furthermore, it was found that $E_F$ can be tuned into
the surface-state gap which in turn may allow the observation of many
interesting topological phenomena, such as the image magnetic monopole
induced by an electric point charge \cite{Qi09S,Zang10PRB}.
Theoretical work on magnetically doped TIs has also been very active
\cite{Zhu11PRL,Biswas10PRB,Liu09PRL,Garate10PRL,Niu11APL}. These
studies predict a variety of effects due to the RKKY interactions
between different magnetic impurities mediated by the surface states.
For example, Zhu et al. \cite{Zhu11PRL} predict that such interaction
consists of Heisenberg-like, Ising-like, and
Dzyaloshinskii-Moriya-like terms.  They devise a new way to control
surface magnetism using electrical field or even by changing the
average spacing between different impurities at the surface, e.g., by
changing their doping level.  However, most importantly this surface
magnetism is predicted to be dramatically different from the bulk due
to the different type of magnetic interactions at the surface.

In this paper we investigate magnetically (Fe) doped \BS, at different
doping levels. Using zero field (ZF) muon spin relaxation ($\mu$SR)
and bulk magnetization measurements, we find that even at low doping
levels, the full volume of the material becomes magnetic at a
relatively high temperature ($\sim 250$K). We identify two magnetic
regimes; the first between $\sim 10-250$K and the second below $\sim
10$K. The nature of the magnetism in both regimes, as well as the
effective size of the magnetic moment per doping Fe, are independent
of the doping level. However, the dynamics due to magnetic
fluctuations increase at lower doping. Moreover, near $\sim 250$K, the
size of local magnetic field at the muon site increases more sharply
as the temperature is decreased for higher Fe doping. These results
indicate that Fe-Fe interactions dominate the magnetism at high
temperatures, while the single Fe ion properties dominate in the low
temperature regime. These studies form a basis for future studies of
the magnetism near the surface using depth resolved low energy $\mu$SR
\cite{Morenzoni94PRL,Prokscha08NIMA}, which may shed light on the
interplay between magnetism and topological surfaces states in this
important class of materials.

\section{Experimental}
Single crystals of \dBS, with $x=0,\ 0.1,\ 0.2$ and $0.3$, were grown
from a melt using the Bridgman method. Corresponding amounts of Bi,
Fe, Se of minimum purity 99.99\% were mixed and sealed in evacuated
silica ampules. The ampules were annealed at 820$^\circ$ C over 10 h
for homogenization. The melt was then cooled down to 640$^\circ$ C at
a rate of 4$^\circ$ C/h and then quenched into cold water.
Well-formed silvery crystal rods were obtained, which could be easily
cleaved into plates with flat shiny surfaces. The magnetization
measurement reported here were obtained using a Physical Properties
Measurement System (PPMS, Quantum Design).
 
The \dBS\ ($x=0.2$, 0.3) crystals were characterized at room
temperature by powder X-ray diffraction (XRD) using a D8 Advance
Bruker AXS diffractometer with Cu K$_\alpha$ radiation. Additionally,
the \dBS\ ($x=0.2$) crystal was studied by means of neutron powder
diffraction (NPD) at the SINQ spallation source at the Paul Scherrer
Institute (PSI, Switzerland), using the high-resolution diffractometer
for thermal neutrons HRPT\cite{Fischer00PB} (wavelengths
$\lambda=1.494$~{\AA} and 1.155~{\AA}). The NPD patterns were measured
at temperatures in the range $2-300$~K. For the diffraction
measurements, part of the crystal was powderized and loaded into a
vanadium container with an indium seal in a He glove box. Refinement
of the crystal structure parameters was done using {\tt FULLPROF}
software \cite{Juan93PB}, with its internal tables for neutron
scattering lengths.

The \msr\ experiments were performed on the DOLLY spectrometer at PSI,
Switzerland. In these experiments $100 \%$ polarized (along the beam
direction, $z$) positive muons are implanted in the sample. In our
measurements single crystals were suspended on an aluminized Mylar
tape and mounted into a He gas flow cryostat. The crystals were
aligned such that the initial polarization of implanted muons is along
the $c$-axis (and the applied field in non-zero field measurements).
Each implanted muon decays (lifetime $\tau_{\mu}=2.2$ $\mu$sec)
emitting a positron preferentially in the direction of its
polarization at the time of decay. Using appropriately positioned
detectors, one measures the asymmetry of the muon beta decay along $z$
as a function of time $A(t)$, which is proportional to the time
evolution of the muon spin polarization. $A(t)$ depends on the
distribution of internal magnetic fields and their temporal
fluctuations.

\section{Results}
\subsection{Magnetization measurements}
\begin{figure}[h]
  \centering
  \includegraphics[width=\columnwidth]{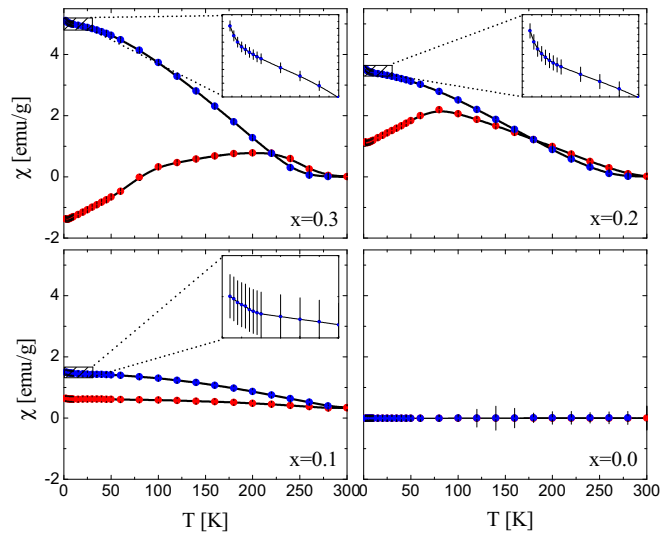}
  \caption{The FC (blue) and ZFC (red) susceptibility measured in
    $B=100$ mT magnetic field ($B \perp c$) as a function of
    temperature for the different Fe doping. The solid lines are a
    guide to the eye and the insets are a zoom of the shaded areas
    ($0-30$~K).}
  \label{MvsT}
\end{figure}

In Fig.~\ref{MvsT} we show the magnetic susceptibility, $\chi$, as a
function of temperature.
In this figure we present both zero field cooled (ZFC) and field
cooled (FC) measurements in a magnetic field of $B=100$ mT applied
perpendicular to the crystallographic c-axis ($B \perp c$). The mother
compound \BS\ gives a small diamagnetic signal with almost no
temperature dependence. In contrast, Fe doping makes \dBS\ magnetic
below $\sim 250$K, confirmed by a clear split between the ZFC and FC
susceptibility. The magnetic nature seems to change at temperatures
below $\sim 10$ K, where a small abrupt upturn is observed for all
doping levels measured (see insets of Fig.~\ref{MvsT}). Similar
behaviour is observed in measurements with the magnetic field applied
parallel to the $c$-axis ($B \parallel c$), but with a more pronounced
upturn below $\sim 10$ and smaller $\chi$ values in general.

\begin{figure}[h]
  \centering
  \includegraphics[width=\columnwidth]{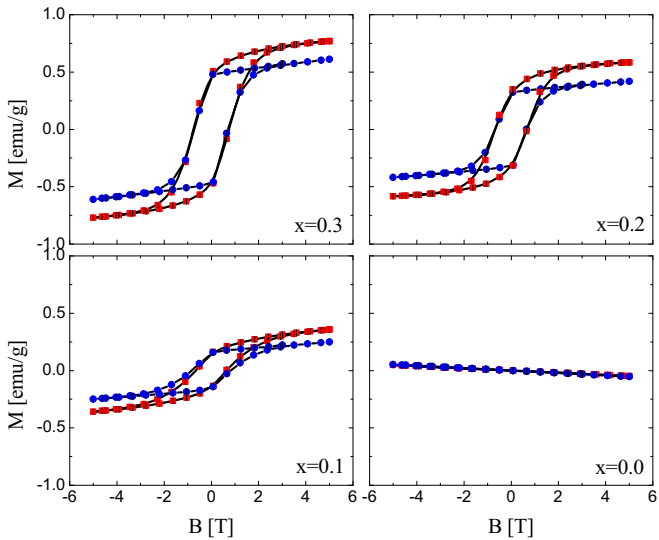}
  \caption{The magnetization as a function of magnetic field ($B \perp
    c$) for the different Fe doping. The red squares and blue circles
    are measurements at $T=2$ and $35$ K, respectively. The solid
    lines are a guide to the eye.}
  \label{MvsB}
\end{figure}

\begin{figure}[h]
  \centering
  \includegraphics[width=\columnwidth]{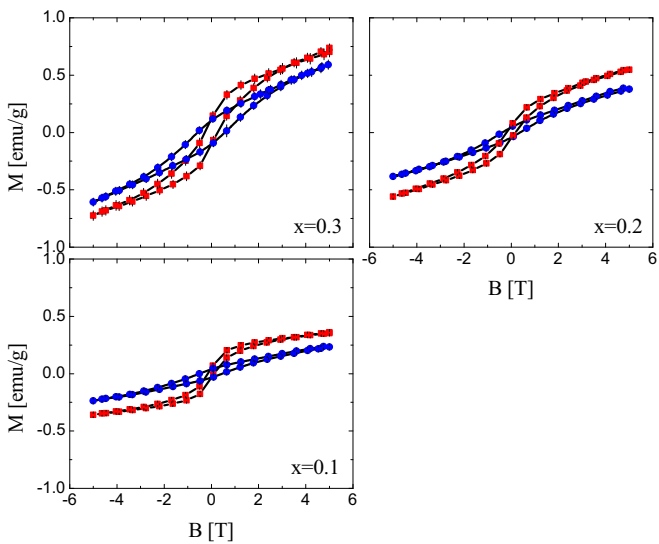}
  \caption{The magnetization as a function of magnetic field ($B
    \parallel c$) for the different Fe doping.  The red squares and
    blue circles are measurements at $T=2$ and $35$ K, respectively.
    The solid lines are a guide to the eye.}
  \label{MvsBparC}
\end{figure}

Measurements of the hysteresis loops in all four samples are shown in
Figs.~\ref{MvsB} ($B \perp c$) and Fig.~\ref{MvsBparC} ($B \parallel
c$).
These were measured at two different temperatures; 2 and 35 K to
better show the difference above and below 10 K. The diamagnetic and
temperature independent nature of the mother compound \BS\ ($x=0$) is
evident in the absence of hysteresis and the negative slope of $M$ vs.
$B$ yielding $\chi \sim -1 \time 10^{-6}$ emu/g. However, for all
doped samples we observe a clear opening of the hysteresis when the
field is applied perpendicular to the $c$-axis (Fig.~\ref{MvsB}). In
contrast, we find that the hysteresis loops are much narrower when the
field is applied along the $c$-axis (Fig.~\ref{MvsBparC}). This strong
anisotropy indicates that the magnetic easy axis in \dBS\ lies within
the Bi layer and almost perpendicular to the $c$-axis.
Furthermore, the hysteresis loops seem to be narrower below 10 K
pointing to the possibility of a different orientation of the magnetic
easy axis between the high and low temperature regimes.

Note, in the $B \perp c$ measurements (Fig.~\ref{MvsB}) the remanence
depends only weakly on temperature, but the saturation of the
magnetization differs significantly. In addition, we find that the
magnetization saturates gradually at 2K, while at 35K it is much
clearer, though lower in magnitude. This difference is again a
possible indication of the different nature of magnetic ordering below
10 K, e.g. due to a slight reorientation of the magnetic easy axis.
Our measurements are similar to those reported earlier\cite{Chen10S}
on similar samples, though the magnetization values measured in our
samples are significantly higher for similar doping levels. As we show
below, XRD, NPD and \msr\ measurements strongly indicate that the
observed magnetization cannot be attributed to ferromagnetic inclusion
of Fe-Se binary compounds.
\begin{table*}[tb]
\begin{tabular}{|l||l|l|l|}
\hline
                        &   XRD (300K)      &  NPD (300K)      & NPD (2K)    \\ \hline \hline
a (\AA)                 & 4.14072(10)         & 4.13914(4)       & 4.12518(3)  \\ \hline
c (\AA)                 &  28.64692(60)       &  28.63151(56)    & 28.47718(40)\\ \hline
Bi $(0,0,z)$              & 0.40057(6)          & 0.40107(14)      & 0.40125(9)  \\ \hline
B$_{\rm Bi}$(\AA$^2$)    & 0.997(52)           & 1.743(67)        & 0.498(45)   \\ \hline
Se2 $(0,0,z)$             & 0.20984(10)         & 0.21071(8)       & 0.21062(6)  \\ \hline
B$_{\rm Se2}$(\AA$^2$)   & 3.173(144)          & 1.393(97)        & 0.745(65)   \\ \hline
%Fe occupancy on Bi-site    & 0.364(20)           & 0.315(215)       &0.105(175)\\ \hline
$R_{p}$,$R_{wp}$,$R_{exp}$ & 6.55, 8.46, 3.77    & 3.80, 4.75, 3.65 &3.46, 4.46, 2.00 \\ \hline
$\chi^{2}$                 & 5.04                & 1.69             & 4.99 \\ \hline
\end{tabular}
\caption{Crystal structure parameters of \dBS\ ($x=0.2$) obtained from XRD (300 K) and NPD (300 and 2 K) refined in the space group R-3m (no. 166) with Se1 in $(3a)$ position $(0,0,0)$, Bi and Fe in $(6c)$ position $(0,0,z)$, and Se2 in $(6c)$ position $(0,0,z)$. B$_{\rm Bi}$ and B$_{\rm Se2}$ are the atomic displacement parameters of Bi and Se2 atoms, respectrively. $R_{p}$,$R_{wp}$ and $R_{exp}$ are the reliability factors and $\chi^{2}$ is the global chi-square (Bragg contribution).} \label{Xtal}
\end{table*}

\subsection{NPD and XRD measurements}
Both room-temperature NPD and XRD experiments show that the samples
consist mainly of the rombohedral \dBS\ phase (space group R-3m).
However, we also detect a small impurity phase of Fe$_{3}$Se$_{4}$
(about 5\% mass), which has a monoclinic structure and orders
ferrimagnetically below $T_c \sim 314$~K \cite{Andresen70ACS}.
Difference Fourier density map analysis reveals that Fe substitutes Bi
atoms in the structure. The refined Fe occupancy is in agreement with
the nominal composition, but the deduced values contain large
systematic errors.  This is due to the presence of the small impurity
phase and a strong preferred orientation of the flat-shape
crystallites in the powder along (001)-direction. This in turn affects
their packing and the resulting diffraction patterns seen in both XRD
and NPD. We account for this artifact by using the March formula in
the {\tt FULLPROF} refinement analysis. Note, since the scattering
lengths of Bi and Fe are quite close (8.53~fm and 9.45~fm,
respectively), NPD is not very sensitive to the Bi/Fe-occupancy ratio.
Therefore, their sum was simply fixed at 2. High statistics NPD data,
which were collected at two different temperatures 2K and 20K, do not
exhibit any statistically significant difference due to possible
magnetic contribution. This may be due to a small magnetic moment on
the Fe ions or due to a disordered nature of the magnetic phase in
these compounds. The crystal structure parameters for the main phase
refined from XRD (300K) and NPD (2K and 300K) are presented in the
Table~\ref{Xtal}.

\subsection{\msr\ measurements}
\begin{figure}[h]
  \centering
  \includegraphics[width=0.9\columnwidth]{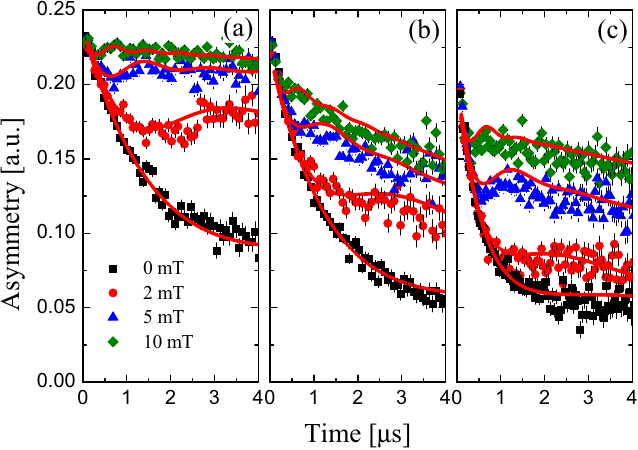}
  \caption{The asymmetry measured in the $x=0.2$ compound with
    different applied longitudinal magnetic fields and temperatures:
    (a) 35K, (b) 5K and (c) 1.6K. The lines are fits to
    Eq.~(\ref{LKTExp}).}
  \label{AsyvsB}
\end{figure}

Compared to conventional dc magnetization measurements, \msr\ is a
much more sensitive method to investigate the magnetism of these
compounds. It also has two very important advantages: (I) detecting
intrinsic magnetism even in zero applied magnetic fields and (II)
since it is a local probe method, the contribution of impurity phases
is proportional to their volume fraction and can be easily
disentangled from the contribution of the studied compound. Typical
\msr\ asymmetries measured in the $x=0.2$ sample are shown in
Fig.~\ref{AsyvsB}.
Similar behaviour was observed in the other two Fe doped compounds.
Note, in zero applied field, the asymmetry relaxes to finite value. A
priori, one cannot determine the source of this relaxation. It can be
due to either a distribution of internal static fields, dynamic
fields, or a combination of the two. However, it is evident in
Fig.~\ref{AsyvsB} that application of an external magnetic field along
the initial direction of the muon spin (so-called longitudinal field
or LF-\msr) decouples the relaxation \cite{muSR2010}. This decoupling,
together with the appearance of ``wiggles'' in the relaxation curves
when the field is applied, are evidence of the quasi-static nature of
the local fields experienced by the implanted muons, i.e., the local
field is predominantly static with a smaller dynamic component. In
this case the relaxation can be described using a phenomenological
function; a Lorentzian Kubo-Toyabe (LKT) multiplied by an exponential
\cite{Salman10PRB}. The LKT accounts for the relaxation due to the
distribution of static fields while the exponential accounts for the
depolarization due to the small dynamic component. The asymmetries in
Fig.~\ref{AsyvsB} were fit to a function of the form
\begin{equation} \label{LKTExp} 
A(t)=A_0 P_{LKT}(B,\Delta,t) e^{-t/T_1},
\end{equation}
where $A_0$ is the initial asymmetry, $P_{LKT}(B,\Delta,t)$ is the
depolarization function due to a Lorentzian distribution of static
fields with width $\Delta$ in the applied field $B$, and $1/T_1$ is
the spin lattice relaxation rate due to the fluctuating transverse
magnetic field component at the muon site. In order to obtain a
reliable fit in Fig.~\ref{AsyvsB}, we assume a temperature and field
independent value of $A_0$ and a field independent (but temperature
dependent) value of $\Delta$, with only $1/T_1$ allowed to vary as a
function of the applied field.  These assumptions produce reliable
fits and very good agreement with the measured asymmetries. Note that
Eq.~(\ref{LKTExp}) account for the {\em full signal}, and therefore no
impurity phase (e.g. Fe$_3$Se$_4$) contribution is detected in our
measurements. This could be due to the small volume fraction of this
phase and its correspondingly small contribution to our measured
signal. Alternatively, the single crystals used for these
measurements, which were cleaved from large crystal rods, are free
from impurities.

\begin{figure}[h]
  \centering
  \includegraphics[width=0.8\columnwidth]{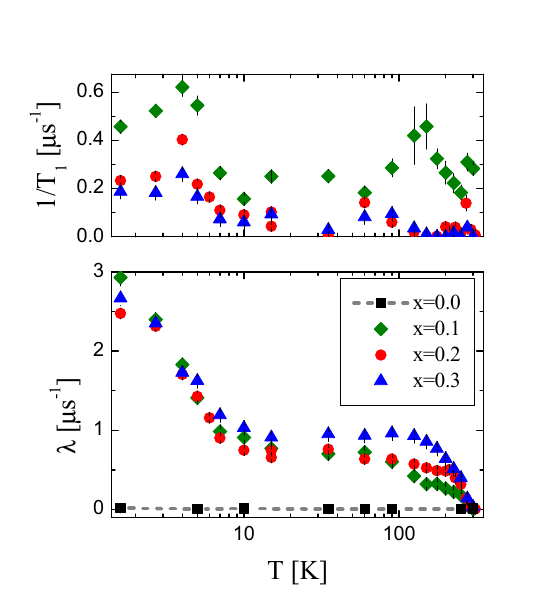}
  \caption{The temperature dependence of $\lambda$ (bottom) and
    $1/T_1$ (top) obtained from the ZF \msr\ data for the different Fe
    doping.}
  \label{LamvsT}
\end{figure}

Typically, the muon spin depolarization function due to a randomly
oriented quasi-static local magnetic field exhibits a dip at early
times, followed by a recovery to $\sim 1/3$ of the initial asymmetry
which continues to relax in an exponential-like manner at longer
times. The absence of a clear dip in the ZF measurements is due to the
relatively small value of $\Delta$ ($\sim 1$MHz) and the dynamic
component of the local field. This makes it difficult to fit reliably
all our data to Eq.~(\ref{LKTExp}). Instead we parametrize the
temperature dependence in all three Fe doped compounds using a
function of the form,
\begin{equation} \label{ExpExp}
  A(t)=A_0\left( \frac{2}{3} e^{-\lambda t} +\frac{1}{3} e^{-t/T_1} \right).
\end{equation}
Here, $\lambda$ is a parameter proportional to $\Delta$ (approximately
$\lambda \sim 2 \Delta$) and quantifies the width of the static field
distribution. From these fits we extract the values of $\lambda$ and
$1/T_1$ as a function of temperature shown in Fig.~\ref{LamvsT} for
the three Fe doped samples.
Note that, in agreement with the magnetization measurements, we
observe a large increase in $\lambda$ below $\sim 250$ and $\sim 10$K.
These coincide with broad and weak peaks in the values of $1/T_1$,
indicating that the Fe doped \BS\ undergoes two different types of
magnetic ordering at these temperatures. However, while $\lambda$
below $\sim 250$~K seems to depend strongly on the doping level, it is
almost independent of $x$ below $\sim 10$ K. In fact, the main
difference between the doping levels appears just below 250 K, where
the increase in $\lambda$ is sharper for higher doping. The spin
lattice relaxation rate, $1/T_1$, depends more strongly on $x$. Its
value increases with decreasing $x$, indicating that the fluctuations
of local fields are faster for lower doping. In all three doping
levels a sharp peak is observed at $\sim 5$ K, which we attribute to
the magnetic phase transition below $\sim 10$ K. In contrast, the peak
at high temperature is broader for all $x$ values and it seems to
shift to lower temperatures for larger Fe doping. We attribute this
peak to the magnetic phase transition occurring below $\sim 250$ K.
The width and quality of the results do not allow accurate
determination of the exact transition temperature, nevertheless, they
indicate a more disordered and possibly inhomogeneous magnetic
transition (typically seen in dilute spin glasses \cite{Keren96PRL}).

In contrast to the doped samples, the muon spin relaxation rate in the
mother compound is very small and has a very weak temperature
dependence. Fitting the relaxation to an exponential function gives a
rate of $\sim 0.008$ $\mu$s$^{-1}$ (bottom panel of
Fig.~\ref{LamvsT}). The only source of relaxation in this case is
expected to be dominated by the Bi nuclear magnetic moments. However,
the observed relaxation is much smaller than one expects from dipolar
fields of nuclear moments and therefore strongly indicates that the
muons stopping sites are far away from Bi. In our case, the muons are
most probably located in the van der Waals gap between layers of \BS.

\section{Discussion}
\begin{figure}[h]
  \centering
  \includegraphics[width=0.8\columnwidth]{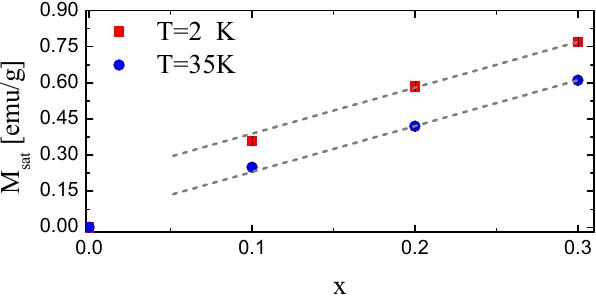}
  \caption{The saturation magnetization as a function of x, the Fe
    doping level at 35K and 2 K. The dashed lines are a linear fit of
    the data.}
  \label{Mvsx}
\end{figure}

We start by discussing the bulk magnetization results. The values of
saturation magnetization, $M_{\rm sat}$, extracted from
Fig.~\ref{MvsB}, are plotted as a function of $x$ in Fig.~\ref{Mvsx}.
For all doping levels we find that $M_{\rm sat}$ is higher at $T=2$ K
than 35 K. Moreover, a close inspection of $M_{\rm sat}$ at both
temperatures shows that it increases linearly with $x$
(Fig.~\ref{Mvsx}). This behaviour indicates that the size of the
magnetic moment of Fe does not depend strongly on $x$ and that the
increase in magnetization simply scales with the number of Fe ions in
the sample.
In fact, from the slope of $M_{\rm sat}$ as a function of $x$ we find
that the contribution of each Fe ion to the magnetization is equal
(within error) at both temperatures and amounts to $1.91 \pm 0.06$
emu/g. From $M_{\rm sat}$ we estimate a magnetic moment per Fe of
0.34(7) and 0.25(4) $\mu_B$ at 2 and 35 K, respectively. Note,
however, this contribution must decrease for much lower doping in
order to approach the non-magnetic $x \rightarrow 0$ limit.

The \msr\ results, summarized in Fig.~\ref{LamvsT}, show that the
width of the static field distribution (or $\lambda$), which is
proportional to the local magnetization, depends only weakly on $x$
below $\sim 50$ K. Since the muon is a local probe it is mostly
sensitive to the dipolar field from its nearest neighbour magnetic
moment. Therefore, the $x$ independent $\lambda$, together with our
previous conclusion regarding the $x$ independent size of Fe magnetic
moment, imply that the implanted muons stop at a similar average
distance from a neighbouring Fe in all samples. However, one expects
that the implanted muons have higher probability to stop closer to a
Fe ion for larger $x$. This contradiction may be explained if the muon
stops within the van der Waals gap in the material, as we concluded
from the small relaxation rates measured in the mother compound. In
such a case the distribution of dipolar fields experienced by the
muons from the Fe ions, which occupy a site in the Bi layer, does not
depend strongly on $x$. The width of this distribution can be
calculated \cite{Salman12PP} by distributing $x/2$ Fe ions randomly on
a monolayer of a triangular lattice (of Bi, lattice constant $a$), and
summing up their contribution on muons which are located $4.76$~{\AA}
away from it (within the van der Waals gap). Although the proposed
model is very simple, we find that for an average Fe moment of
0.34$\mu_B$ we obtain values of $\lambda$ which are consistent with
the experimental results.

The measurements presented here indicate that the low and high
temperature magnetic phases differ in the underlying interactions.
The high temperature regime seems to be associated with Fe-Fe
coupling, and therefore depends on $x$ which affects the average
distance between Fe ions. In contrast, the weak $x$ dependence at low
temperature is strong evidence that the magnetization is dominated by
the Fe single ion properties and its interaction with the \BS\
lattice, e.g. crystal field etc. Hence, as the temperature is
decreased the smaller single ion interaction with the lattice becomes
more relevant, altering the high temperature size and ordering of the
Fe moments. This is consistent with the strong anisotropy detected in
magnetization measurement, which show a strong tendency of the Fe
moments to align perpendicular to the $c$-axis, especially at low
temperatures. Moreover, considering the low density of Fe ions, and
therefore the fairly long range (a few $a$) magnetic Fe-Fe
interaction, it is most probably mediated by charge carriers which
extend over these length scales. Such effects have been observed in Fe
doped Bi$_2$Te$_3$, though with magnetic transition at much lower
temperatures ($<15$ K) \cite{Kulbachinskii01JEATPL} which also depend
strongly on $x$. This was attributed to donor type Fe doping in both
Bi$_2$Te$_3$ and \BS. However, in that study it was found that \dBS\
with $x=0.04$ is non-magnetic. Furthermore, that the magnetic
transition temperatures in Bi$_{2-x}$Fe$_x$Te$_3$ depends strongly on
$x$ and that the easy axis is along $c$. Finally, it is important to
point out here that recently Lasia et al.\cite{Lasia12condmat}
predicted that in a magnetically doped TI the magnetic moments order
ferromagnetically at zero tempertature, but udergo a first order
transition to a spin density wave (SDW) at higher temperature before
they reach the paramagnetic phase. Our results may be consistent with
this model, though it is unlikely that the high temperature phase,
which can be detected clearly in dc magnetization measurements, is due
to a SDW ordering. Such ordering is typically associated with small
bulk magnetization.

% \begin{table}
%   \begin{tabular}{|l|l|l|l|} \hline
%     & $\Delta(T=1.6 {\rm K})$ & $\Delta(T=5 {\rm K})$& $\Delta(T=35 {\rm K})$ \\ \hline
%     $x=0.1$ & $1.04(4)$ MHz & & \\ \hline
%     $x=0.2$ & $1.00(3)$ MHz& $0.41(1)$ MHz&$0.34(1)$ MHz\\ \hline
%     $x=0.3$ & $1.11(4)$ MHz& & $0.560(16)$ MHz \\ \hline
% \end{tabular}
% \end{table}

\section{Summary and Conclusions}
In conclusion, we find that even at Fe doping levels as low as 5\%,
the full volume of the \BS\ becomes magnetic at a relatively high
temperature of $\sim 250$K. We identify two magnetic regimes; the
first between $\sim 10-250$K and the second below $\sim 10$K. Bulk
magnetization measurements show that the average size of Fe magnetic
moment depends on temperatures but is almost independent of doping,
and that the at the Fe moments align perpendicular to the
crystallographic $c$-axis. Our \msr\ results indicate that the low
temperature regime is independent of doping level and is more likely
determined by Fe single ion properties and crystal field effects. In
contrast, the nature of the magnetism in the high temperature regime
differs, such that Fe-Fe interactions become more important, and
therefore the local magnetization depends more on doping. The observed
magnetism is homogeneous in the full volume of the sample and cannot
be attributed to magnetic impurities. Finally, we conclude that the
appearance of local static magnetic field at high temperatures and low
doping is strong evidence that the Fe-Fe interactions are dominated by
conduction electrons, e.g. RKKY type interaction. Therefore, near the
surface, we expect that the magnetization will be affected differently
by the topological surface states in the two temperature regimes. At
low temperatures, the broken crystal symmetry will affect crystal
field, while the change in the nature of conduction carriers due to
topological surface states will affect the high temperature magnetism.
Low energy \msr\ measurements are underway to explore the magnetism
and its dependence on surface proximity. The results presented here
will provide an excellent point of reference to detect any changes in
the nature of the magnetism near the surface of Fe doped \BS.

\bibliographystyle{apsrev}
%\bibliography{/home/l_salman/LaTex/references}
\newcommand{\noopsort}[1]{} \newcommand{\printfirst}[2]{#1}
  \newcommand{\singleletter}[1]{#1} \newcommand{\switchargs}[2]{#2#1}

\end{document}